\documentclass[preprint,aps,prd,amsmath,amssymb]{revtex4}
\pdfoutput=1
\usepackage{graphicx}
\usepackage{color}
\newcommand{\beq}{\begin{eqnarray}}
\newcommand{\eeq}{\end{eqnarray}}

\newcommand{\bmp}{\noindent\begin{minipage}{16cm}}
\newcommand{\emp}{\end{minipage}\vskip 7mm} 

\usepackage{dcolumn}
\usepackage{bm}
\usepackage{bbm}
\usepackage{subfigure}
\usepackage{pxfonts}

\usepackage{epsfig}

\def\lsim{\mathrel{\rlap{\lower4pt\hbox{\hskip1pt$\sim$}}
    \raise1pt\hbox{$<$}}}                
\def\gsim{\mathrel{\rlap{\lower4pt\hbox{\hskip1pt$\sim$}}
    \raise1pt\hbox{$>$}}}                

\newcommand{\Slash}[1]{\ooalign{\hfil/\hfil\crcr$#1$}}
\newcommand{\crit}{\rm crit}
\newcommand{\SxSB}{S$\chi$SB }
\newcommand{\gNJLcritline}{
\begin{picture}(150,150)(-15,-15)
   \thinlines
  \put(140,3){$\alpha$}
  \put(3,140){$g$}
  \put(-10,0){\line(1,0){150}}
  \put(0,-10){\line(0,1){150}}
  \thicklines
  \put(100,-10){\line(0,1){35}}
  \put(107,-15){$\alpha^0_{\crit}$}
  \put(35,90){$g_{\crit} = \dfrac{1}{4}\left( 1 + \sqrt{1-\dfrac{\alpha}{\alpha^0_{\crit}}} \right)^2$}
  \put(-2,100){\line(1,0){5}}
  \put(-12,98){$1$}
  \multiput(0,25)(5,0){20}{\line(1,0){2.5}}
  \put(-12,23){$\dfrac{1}{4}$}
  \bezier{300}(0,100)(99.7,50)(99.7,25)
  \put(25,40){$\text{Sym.}$}
  \put(85,60){\SxSB}
\end{picture}}

\begin{document}
\title{\Large  \color{red} Conformal Window of Gauge Theories \\ with \\ Four-Fermion Interactions 
and Ideal Walking}
\author{Hidenori S. Fukano$^{\color{blue}{\clubsuit}}$}
\email{fukano@kmi.nagoya-u.ac.jp} 
\author{Francesco {\sc Sannino}$^{\color{blue}{\varheartsuit}}$}
\email{sannino@cp3.sdu.dk} 
\affiliation{$^{\color{blue}{\clubsuit}}$ Kobayashi-Maskawa Institute for the Origin of Particles and the Universe, \\
Nagoya Universtiy, Nagoya 464-8602, Japan \\~\\
$^{\color{blue}{\varheartsuit}}$ \mbox{{ CP}$^{ \bf 3}${-Origins}, 
Campusvej 55, DK-5230 Odense M, Denmark}}
\begin{flushright}
{\it CP$^3$- Origins-2010-19}
\end{flushright}
\begin{abstract}
We investigate the effects of four-fermion interactions on the phase diagram of strongly interacting theories for any representation as function of the number of colors and flavors. We show that the conformal window, for any representation, shrinks with respect to the case in which the four-fermion interactions are neglected. The anomalous dimension of the mass increases beyond the unity value at the lower boundary of the new conformal window. We plot the new phase diagram which can be used, together with the information about the anomalous dimension,  to propose ideal models of walking technicolor. We discover that when the extended technicolor sector, responsible for giving masses to the standard model fermions, is sufficiently strongly coupled the technicolor theory, in isolation,  must  have an infrared fixed point for the full model to be phenomenologically viable. Using the new phase diagram we show that the simplest one family and minimal walking technicolor models are the archetypes of models of dynamical electroweak symmetry breaking. Our predictions can be verified via first principle lattice simulations.
\end{abstract}

\maketitle

\tableofcontents
\newpage

\section{Introduction}

Models of dynamical breaking of the electroweak symmetry \cite{Weinberg:1979bn,Susskind:1978ms} are theoretically appealing and constitute one of the best motivated extensions of the standard model (SM).  We have proposed several  models \cite{Sannino:2004qp, Hong:2004td,Dietrich:2005wk,Dietrich:2005jn,Gudnason:2006mk,Ryttov:2008xe,Frandsen:2009fs,Frandsen:2009mi,Antipin:2009ks} possessing interesting dynamics relevant for collider phenomenology \cite{Foadi:2007ue,Belyaev:2008yj,Antola:2009wq,Antipin:2010it} and cosmology \cite{Nussinov:1985xr,Barr:1990ca,Bagnasco:1993st,Gudnason:2006ug,Gudnason:2006yj,Kainulainen:2006wq,Kouvaris:2007iq,Kouvaris:2007ay,Khlopov:2007ic,Khlopov:2008ty,Kouvaris:2008hc,Belotsky:2008vh,Cline:2008hr,Nardi:2008ix,Foadi:2008qv,Jarvinen:2009wr,Frandsen:2009mi,Jarvinen:2009mh,Kainulainen:2009rb,Kainulainen:2010pk,Frandsen:2010yj,Kouvaris:2010vv} \footnote{It is worth to note that if dark matter arises as a pseudo Goldstone boson in technicolor models (the TIMP) \cite{Gudnason:2006ug,Gudnason:2006yj,Nardi:2008ix,Foadi:2008qv,Jarvinen:2009wr,Frandsen:2009mi,Jarvinen:2009mh} they should not be confused  with the scaled up version of the ordinary baryon envisioned earlier \cite{Nussinov:1985xr,Barr:1990ca,Bagnasco:1993st}. {}For example one striking feature is that the pseudo Goldstone Bosons can be sufficiently light to be produced at the Large Hadron Collider experiment at CERN.}. The structure of one of these models, known as Minimal Walking Technicolor, has even led to the construction of a new supersymmetric extension of the SM featuring the maximal amount of supersymmetry in four dimension with a clear connection to string theory, i.e. Minimal Super Conformal Technicolor \cite{Antola:2010nt}. These models are also being investigated via first principle lattice simulations \cite{Catterall:2007yx,Catterall:2008qk,DelDebbio:2008zf,Hietanen:2008vc,Hietanen:2009az,Pica:2009hc,Catterall:2009sb,Lucini:2009an,Bursa:2009we,DelDebbio:2010hu,DelDebbio:2010hx,DeGrand:2009hu,DeGrand:2008kx,DeGrand:2009mt,Fodor:2008hm,Fodor:2009ar,Fodor:2009nh,Kogut:2010cz} \footnote{Earlier interesting models \cite{Appelquist:2002me,Appelquist:2003uu,Appelquist:2003hn} have contributed triggering the lattice investigations for the conformal
window with theories featuring fermions in the fundamental representation, see Refs.÷\cite{Appelquist:2009ty,Appelquist:2009ka,Fodor:2009wk,Fodor:2008hn,
Deuzeman:2009mh, Fodor:2009rb,Fodor:2009ff}.}. An up-to-date review is Ref. \cite{Sannino:2009za} while an excellent review updated till 2003 is Ref. \cite{Hill:2002ap}.  

These are also among the most challenging models to work with since they require deep knowledge of gauge dynamics in a regime where perturbation theory fails. In particular, it is of utmost importance to gain information on the nonperturbative dynamics of non-abelian four dimensional gauge theories.  The phase diagram of $SU(N)$ gauge theories as functions of number of flavors, colors and matter representation has been investigated in \cite{Sannino:2004qp,Dietrich:2006cm,Ryttov:2007sr,Ryttov:2007cx,Sannino:2008ha}.  The analytical tools which  have been used here for such an exploration were: i) The conjectured {\it physical} all orders beta function for nonsupersymmetric gauge theories with fermionic matter in arbitrary representations of the gauge group \cite{Ryttov:2007cx}; ii) The truncated Schwinger-Dyson equation (SD) \cite{Appelquist:1988yc,Cohen:1988sq,Miransky:1996pd} (referred also as the ladder approximation in the literature);  The Appelquist-Cohen-Schmaltz (ACS) conjecture \cite{Appelquist:1999hr} which makes use of the counting of the thermal degrees of freedom at high and low temperature. However several very interesting and competing analytic approaches \cite{Grunberg:2000ap,Gardi:1998ch,Grunberg:1996hu,Gies:2005as,Braun:2005uj, Poppitz:2009uq,Poppitz:2009tw,Antipin:2009wr,Antipin:2009dz,Jarvinen:2009fe, Braun:2009ns,Alanen:2009na} have been proposed in the literature. What is interesting is that despite the very different starting points  the various methods agree qualitatively on the main features of the various conformal windows summarized in \cite{Sannino:2009za}. 

Here we would like to reconsider the conformal window obtained using the SD results by adding the effects of four-fermion interactions. The resulting model is the time-honored {\it gauged NJL} model \cite{Bardeen:1985sm,Leung:1985sn} generalized to different matter representations. The main reason for such an investigation is that these type of interactions naturally arise, as effective operators, at the electroweak scale when augmenting the technicolor model with Extended Technicolor Interactions (ETC) \cite{Dimopoulos:1979es,Eichten:1979ah}. These interactions are needed to endow the standard model fermions with a mass when taking the point of view that no fundamental scalars exist in nature. In Refs.÷\cite{Appelquist:2002me,Appelquist:2003uu,Appelquist:2003hn,Fukano:2008iv,Ryttov:2010kc} the reader will find more details about recent progress about the ETC models while in \cite{Sannino:2009za,Poppitz:2009tw}  will find the phase diagram of phenomenologically relevant chiral gauge theories which is known to play an important role for the construction of the ETC models. 

We first introduce and briefly review what is known about the gauged NJL model. We then generalize it to the case of any representation of the underlying gauge group and then investigate the effects of four-fermion interactions on the phase diagram of strongly interacting theories as function of the number of colors and flavors. We show that the conformal window, for any representation, shrinks with respect to the case in which the four-fermion interactions are not considered. We also see that along the lower boundary of the new conformal window the anomalous dimension of the fermion mass is larger than one. 

 We plot the new phase diagram and then use it, together with the information about the anomalous dimension,  to suggest ideal models of (near conformal) walking technicolor. Moreover,  we find that when the ETC sector is sufficiently strongly coupled the technicolor theory, in isolation,  must have developed an infrared fixed point for the full model to be near conformal, i.e. walking. Interestingly, earlier models of one family and minimal walking technicolor constitute ideal extensions of the standard model of particle interactions. The new conformal windows, for a generic number of flavors, colors and matter representation, can be tested via first principle lattice simulations.

\section{Gauged NJL model: A brief review.}
\begin{figure}[htbp]
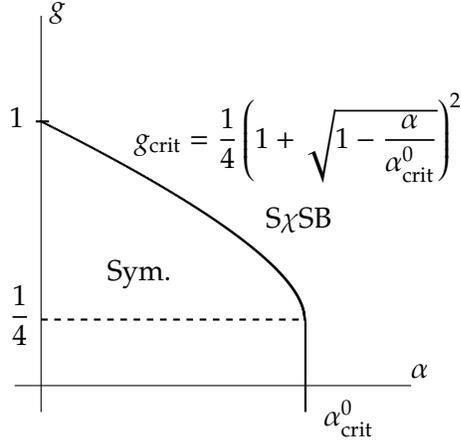

\begin{center}
\gNJLcritline \\
\caption{
 NJL model critical line in the $(\alpha,g)$ plane. It is assumed to separate the chiral spontaneously broken (\SxSB) phase, which is the region above the line, from the  unbroken one (Sym.).\cite{Yamawaki:1996vr}
\label{gNJLcritline}}
\end{center}
\end{figure}

Let's consider an $SU(N)$ gauge theory in which we add four-fermion interactions. This model is known as the gauged NJL model and 
we follow here the nice review by Yamawaki \cite{Yamawaki:1996vr}.
The Lagrangian reads:
\beq
{\cal L} = \bar{\psi} i \Slash{D} \psi +\frac{G}{N_f d\left[\rm r \right]} \left[ (\bar{\psi} \psi)^2 + (\bar{\psi} i \gamma_5 T^a \psi)^2 \right] - \frac{1}{4} \sum^{N^2-1}_{a=1}F^a_{\mu \nu}F^{a\mu\nu}\,.
\eeq
Here $D$ is the standard covariant derivative for the $SU(N)$ gauge theory acting on $N_f$ Dirac fermions in the representation $r$ of the gauge group and $d\left[r \right]$ is its dimension.  $G$ is the four-fermion coupling.  It is convenient to introduce a dimensionless coupling $g =G\Lambda^2/(4\pi^2)$  with $\Lambda$ the cut-off energy scale up to which the gauged NJL model is defined. Using the ladder/rainbow Schwinger-Dyson equation (SD) one arrives at the diagram  of Fig.\ref{gNJLcritline} in the  $(\alpha,g)$ plane. Below the solid line chiral symmetry is intact whilst above it chiral symmetry is spontaneously broken (\SxSB). We recall that the value of $\alpha^0_{\crit} = \pi/(3 C_2(r))$ is the critical one for an $SU(N)$ gauge theory without the four-fermion interactions.
\vskip .5cm 
\noindent 

\section{The Phase Diagram including four-fermion interactions}
Any technicolor model must feature another sector enabling the standard model fermions to acquire a mass term. The simplest models of this type lead to the addition to the technicolor sector, at low energies, of four-fermion interaction. We will show that the net effect is a modification of the conformal window lower boundary.  We will find the relevant result that the presence of four-fermion interactions, de facto, reduces the conformal window area. This fact has an important impact on technicolor extensions of the standard model featuring a traditional ETC sector. Our results show that it is important to study lattice gauge theories including also the effects of the four-fermion interactions. 

To determine the effects of the inclusion of the new operator on the conformal window we start with recalling that the analytical expression for the critical line in the $(\alpha, g)$-plane, shown in Fig.\ref{gNJLcritline}, is given by \cite{Kondo:1988qd,Appelquist:1988fm}
\beq
g_{\rm crit} =
\begin{cases}
\dfrac{1}{4}\left( 1 + \sqrt{1-\dfrac{\alpha}{\alpha^0_{\crit}}} \right)^2 & \text{for } \, 0 < \alpha < \alpha^0_{\crit}\,, 
\\[5ex]
\dfrac{1}{4} & \text{for } \, \alpha = \alpha^0_{\rm crit}\,.
\end{cases}
\label{critical}
\eeq
{}For our purposes it is more convenient to think about fixing $g$ on the critical line with $0 < \alpha < \alpha^0_{\crit}$.  This leads to an associated new critical gauge coupling, which is a monotonically decreasing function of  $g$ on this line, with its maximum value $\alpha^0_{\crit}$:
\beq
\alpha_{\crit} (g_{\rm crit}) = \alpha^0_{\crit} \times 4 \left( \sqrt{g_{\crit}} - g_{\crit}\right) \, \text{ for }\, \frac{1}{4} < g_{\crit} < 1\,.
\eeq
More generally inverting \eqref{critical} we obtain the critical gauge coupling constant for the gauged NJL model  
\beq
\alpha_{\crit} (g)= 
\begin{cases}
4 \left( \sqrt{g} - g \right) \times \alpha^0_{\crit}  & \text{for } \, \dfrac{1}{4} < g < 1\,, 
\\[5ex]
\alpha^0_{\crit} & \text{for } \, 0 < g <\dfrac{1}{4}\,.
\end{cases}
\label{alphacrit-gNJL}
\eeq
We dropped the subscript {\it crit} for $g$. To make contact with the number of flavors, colors and matter representation we compare the critical coupling in Eq.(\ref{alphacrit-gNJL}) with the zero of the full beta function of the underlying gauge theory. We use the universal two-loop beta function for an $SU(N)$ gauge theory with $N_f$ flavors transforming according to the $r$-representation and given by \cite{Caswell:1974gg,Gross:1973ju}
\beq
\beta^{(2)}(\alpha) \equiv - b \alpha^2(\mu) - c \alpha^3(\mu) \,,
\label{2loop-beta}
\eeq
where
\beq
b = {{11 N - 4 N_f {C(r)}}\over{6 \pi}} \,\,,\,\,
c = {{34 N^2 - 2 N_f {C(r)} 
\left[10 N+ 6 {C_2(r)}\right]}\over {24  \pi^2}}\,,
\label{2loop-coef}
\eeq
with $C(r)$ and $C_2(r)$ the Casimirs for the $r$-representation. We have chosen this approximation of the beta function to make direct contact with the traditional approach presented in \cite{Dietrich:2006cm}. 

The asymptotically free condition is: 
\beq
b > 0 \Longleftrightarrow N_f < N^{\text I} \equiv \frac{11N}{4C(r)} \,.
\label{AF-cond}
\eeq
 In addition a zero of this beta function appears when $N_f$ satisfies
\beq
c < 0 \Longleftrightarrow 
N_f > N^{\text{II}}_f 
    \equiv \frac{17 N^2}{C(r) \left[ 10 N+ 6 {C_2(r)} \right]}\,, 
\label{BZ-IRFP-cond}
\eeq
and the associated value of the coupling at the zero reads  
\beq
\alpha_*(N,N_f) &=& - \frac{b}{c} \nonumber\\
                &=& -4\pi
                     \frac{11 N - 4 N_f C(r)}{34 N^2 - 2 N_f C(r) \left[10 N+ 6 {C_2(r)}\right]}\,.
\label{BZ-IRFP-*}
\eeq
To estimate the critical number of flavors  $N^{\crit}_f$, for any color and fixed $g$ in the range $1/4 < g < 1$ , we compare $\alpha_*({N,N_f})$ with $\alpha_{\crit}(g)$
\beq
\alpha_*({N,N^{\crit}_f}) = \alpha_{\crit}(g)\,, 
\eeq
which yields:
\beq
N^{\crit}_f(N,g) = 
\frac{34 N (\sqrt{g} - g) + 33 C_2(r)}{20 N (\sqrt{g}-g) + 12[1 + (\sqrt{g} - g)] C_2(r)}  \cdot \frac{N}{C(r)}\,.
\label{critf-gNJL}
\eeq
In the range  $0 < g < 1/4$, $N^{\crit}_f$ we obtain instead: 
 \beq
N^{\crit}_f(N,g) = 
\frac{17 N  + 66 C_2(r)}{10 N  + 30 C_2(r)} \cdot \frac{N}{C(r)}\,,
\label{critf-g}
\eeq
which is the value estimated in \cite{Dietrich:2006cm} when the four-fermion interaction is neglected. 

\begin{figure}[htbp]
\begin{center}
\includegraphics[width=0.3\textwidth,height=0.3\textwidth]{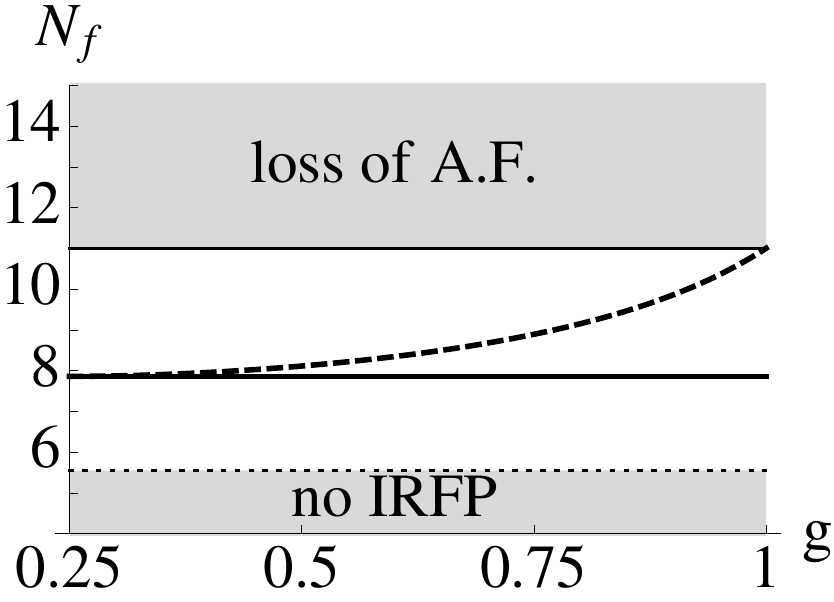} \hspace*{1ex}
\includegraphics[width=0.3\textwidth,height=0.3\textwidth]{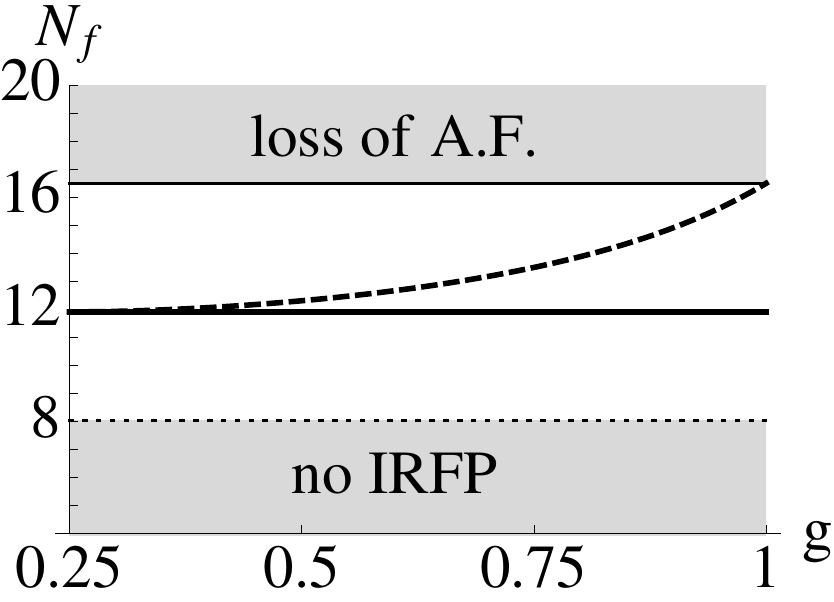} \hspace*{1ex}
\includegraphics[width=0.3\textwidth,height=0.3\textwidth]{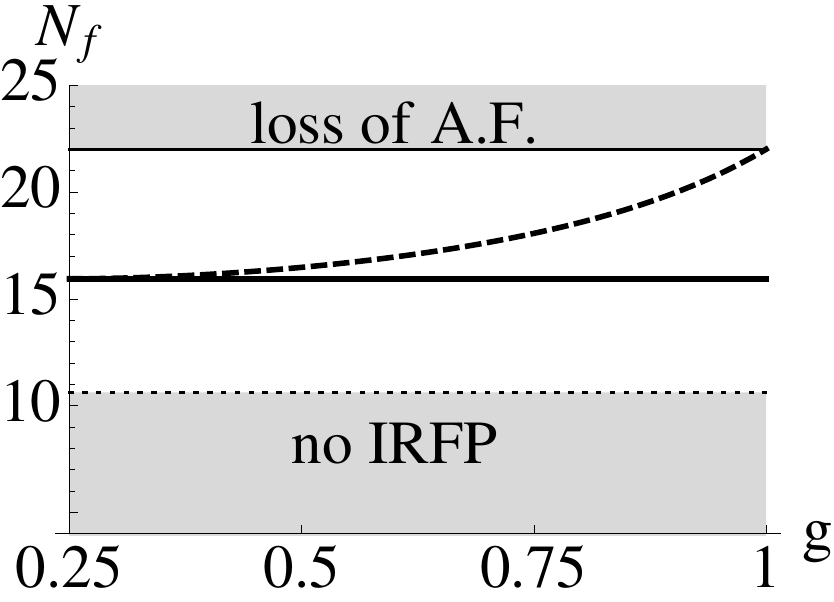} 
\caption{Effect of the four-fermion interactions on the lower bound of the conformal window of $SU(N)$ gauge theories with Dirac fermions transforming according to the fundamental representation : From left to right these figures correspond to $N = 2,3,4$. }
\label{fund}
\end{center}
\end{figure}

\begin{figure}[htbp]
\begin{center}
\includegraphics[width=0.3\textwidth,height=0.3\textwidth]{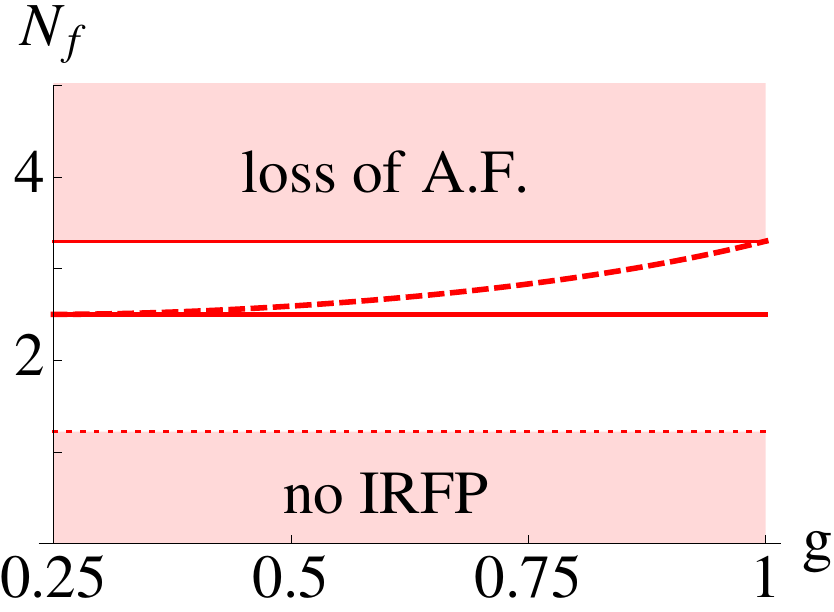} \hspace*{1ex}
\includegraphics[width=0.3\textwidth,height=0.3\textwidth]{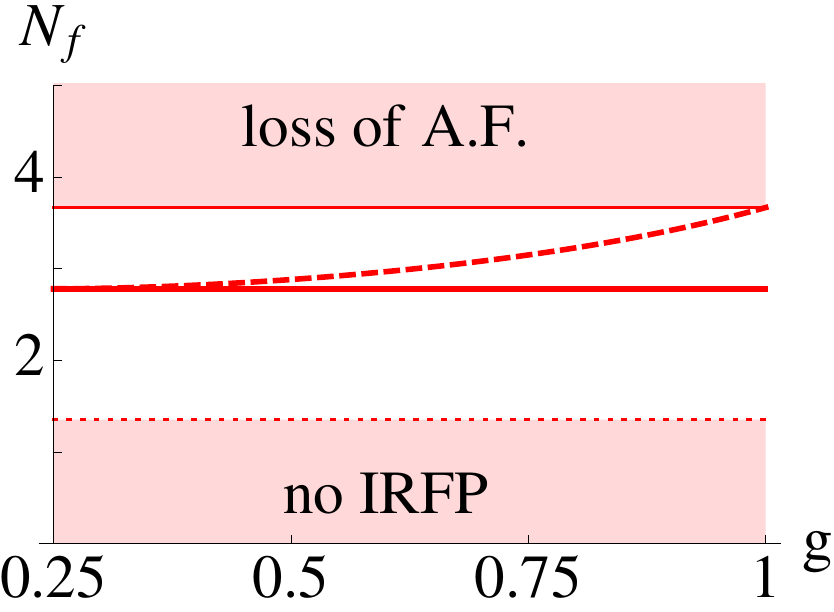} 
\caption{Effect of the four-fermion interactions on the lower bound of the conformal window of $SU(N)$ gauge theories with Dirac fermions transforming according to the two index symmetric representation: From left to right these figures correspond to $N = 3,4$.}
\label{2symm}
\end{center}
\end{figure}

\begin{figure}[htbp]
\begin{center}
\includegraphics[width=0.3\textwidth,height=0.3\textwidth]{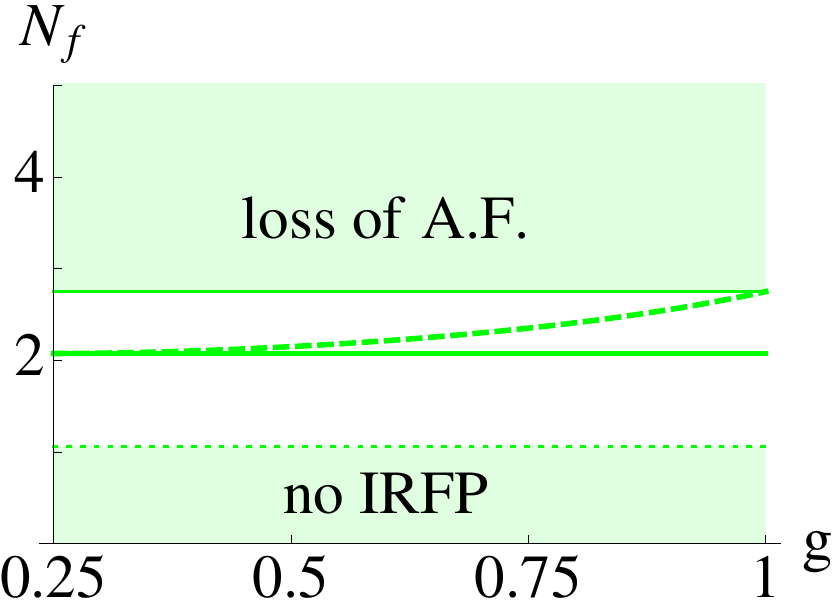} \hspace*{1ex}
\includegraphics[width=0.3\textwidth,height=0.3\textwidth]{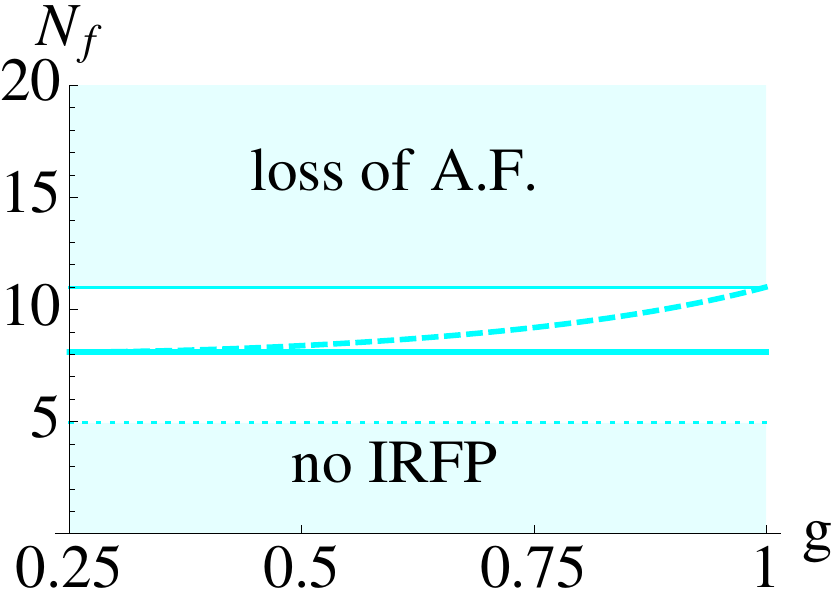}
\caption{Effect of the four-fermion interactions on the lower bound of the conformal window of $SU(N)$ gauge theories with Dirac fermions transforming according to the adjoint and two index anti-symmetric representation: From left to right these figures correspond to the adjoint for $N = 2,\cdots$ and two index anti-symetric for $N = 4$.}
\label{2asymm-adj}
\end{center}
\end{figure}

The general result, as announced at the beginning of this section, is that the area of the conformal window substantially reduces compared to the case of the gauge theory alone. To better elucidate this point we plot  $N^{\crit}_f$ as function of the coupling $g$ for: i) the fundamental representation in Fig.\ref{fund} (the first plot from the left corresponds to $SU(2)$, the middle one to $SU(3)$ and the one on the right to $SU(4)$); ii) the two index symmetric representation is shown in Fig.~\ref{2symm} (the left plot corresponds to $SU(3)$ and the other to $SU(4)$); iii)  finally  in Fig.~\ref{2asymm-adj} the first plot from the left corresponds to $SU(2)$ with matter in the adjoint representation while the other one to two index antisymmetric for $SU(4)$.  The  upper solid lines for all these figures corresponds to the loss of asymptotic freedom encoded in $N^{\text{I}}_f$; the dotted lines correspond to when the two-loops beta function no longer supports a zero and it is encoded in $N^{\text{II}}_f$; the lower solid lines is the SD result for $N^{\crit}_f$ given in Eq.(\ref{critf-g}) when neglecting the four-fermion interactions and finally the dashed curve is $N^{\crit}_f (N,g)$ given in Eq.(\ref{critf-gNJL}) for the gauged NJL model. 
 
\section{Large Anomalous Dimensions}
\begin{figure}[htbp]
\begin{center}
\includegraphics[width=0.3\textwidth,height=0.3\textwidth]{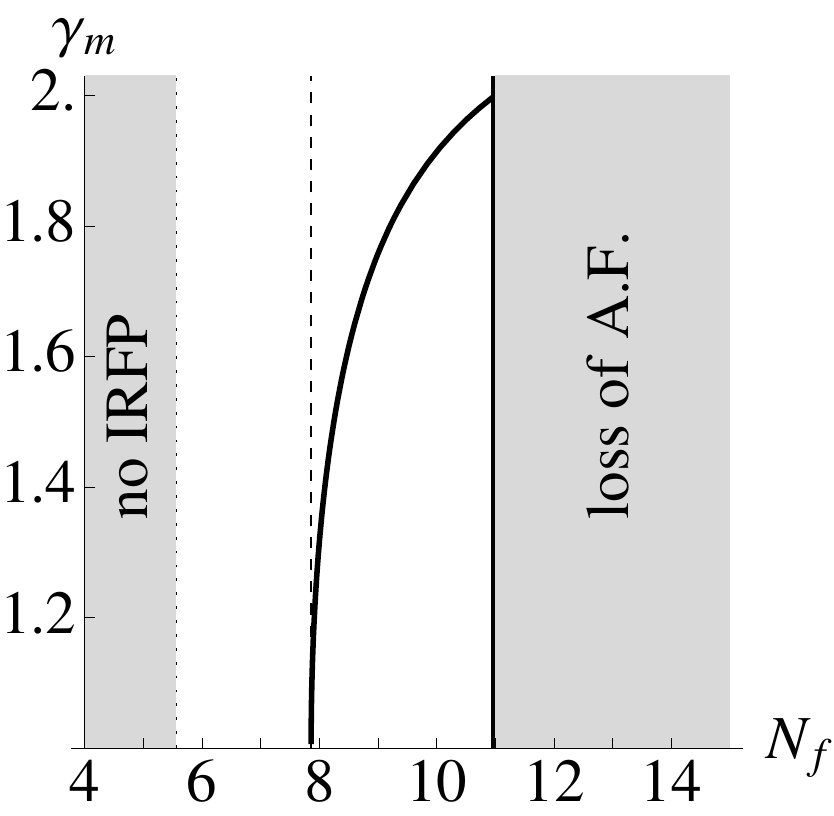} \hspace*{1ex}
\includegraphics[width=0.3\textwidth,height=0.3\textwidth]{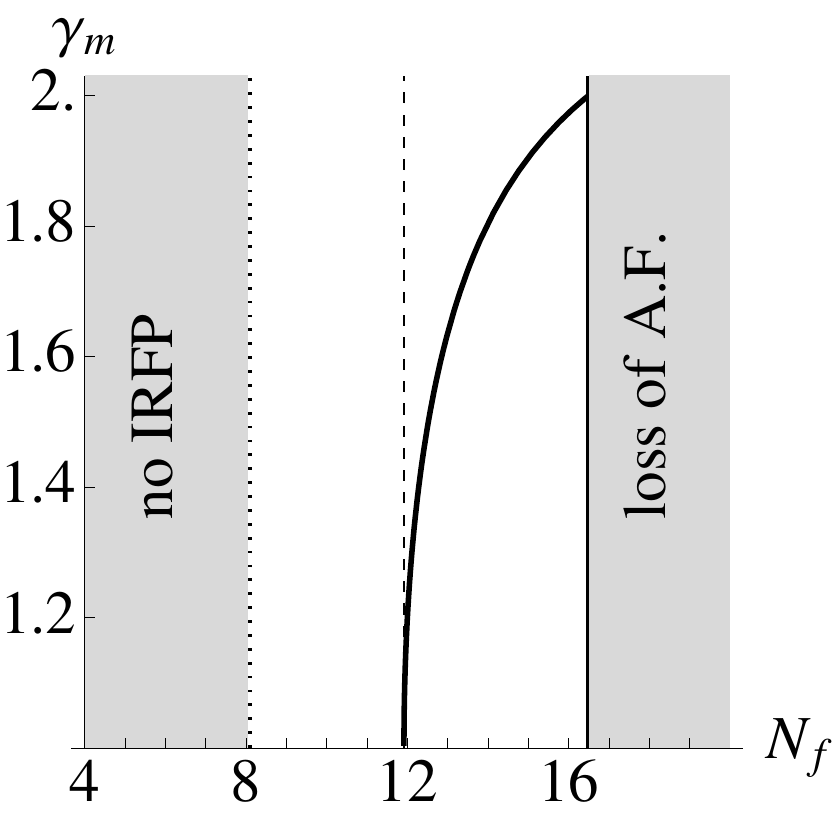} \hspace*{1ex}
\includegraphics[width=0.3\textwidth,height=0.3\textwidth]{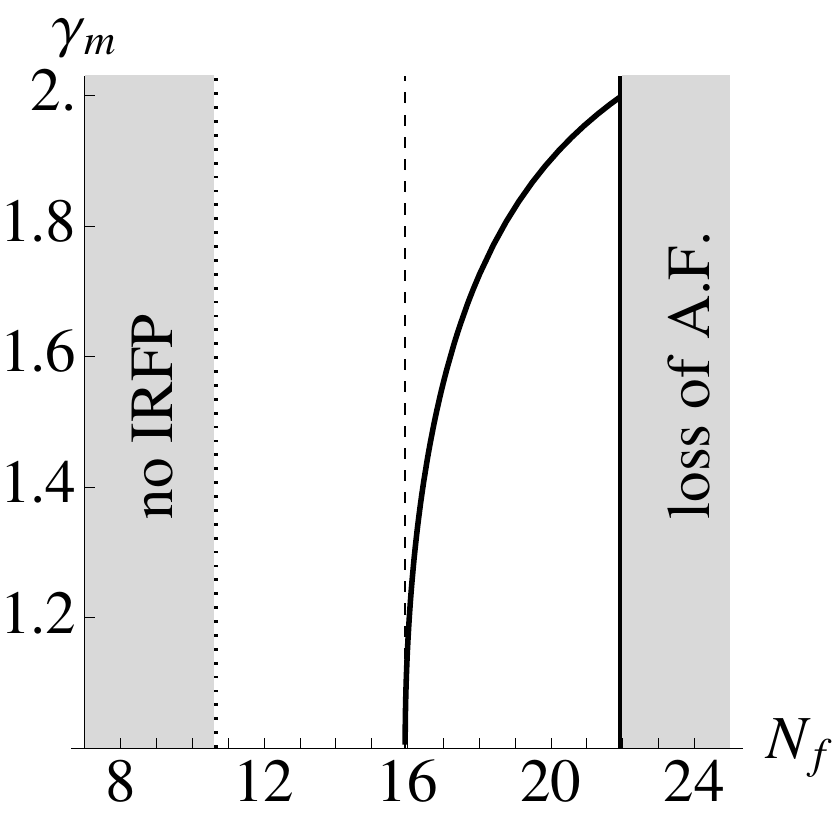} 
\caption{Effect of the four-fermion interactions on the anomalous dimension on the critical line for an $SU(N)$ gauge theories with Dirac fermions transforming according to the two fundamental representation: From left to right these figures correspond to $N = 2,3,4$. }
\label{gamma-fund}
\end{center}
\end{figure}

\begin{figure}[htbp]
\begin{center}
\includegraphics[width=0.3\textwidth,height=0.3\textwidth]{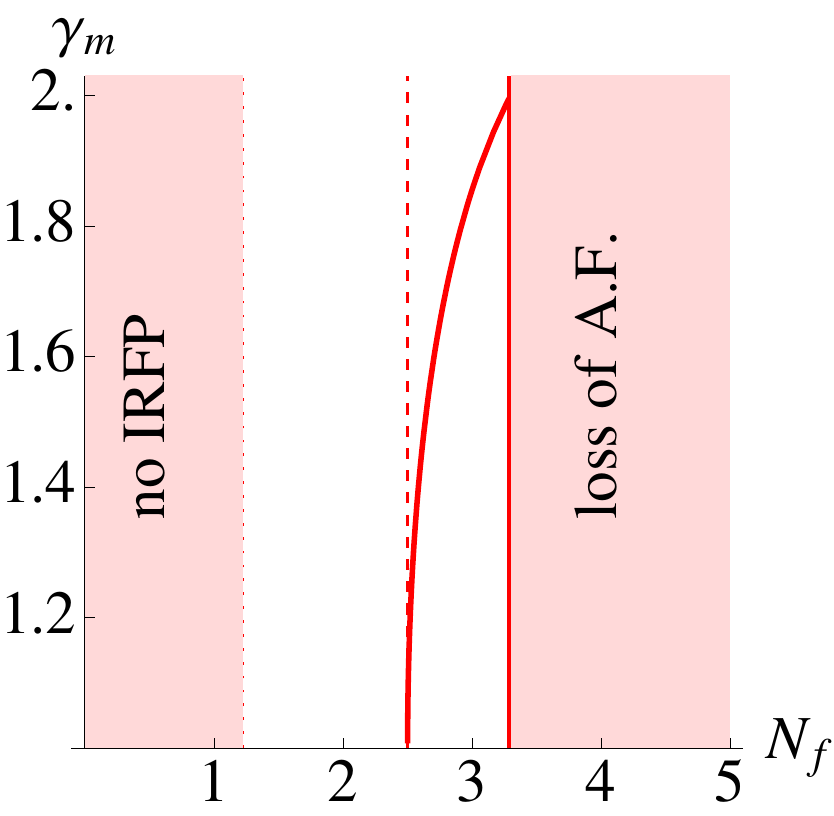} \hspace*{1ex}
\includegraphics[width=0.3\textwidth,height=0.3\textwidth]{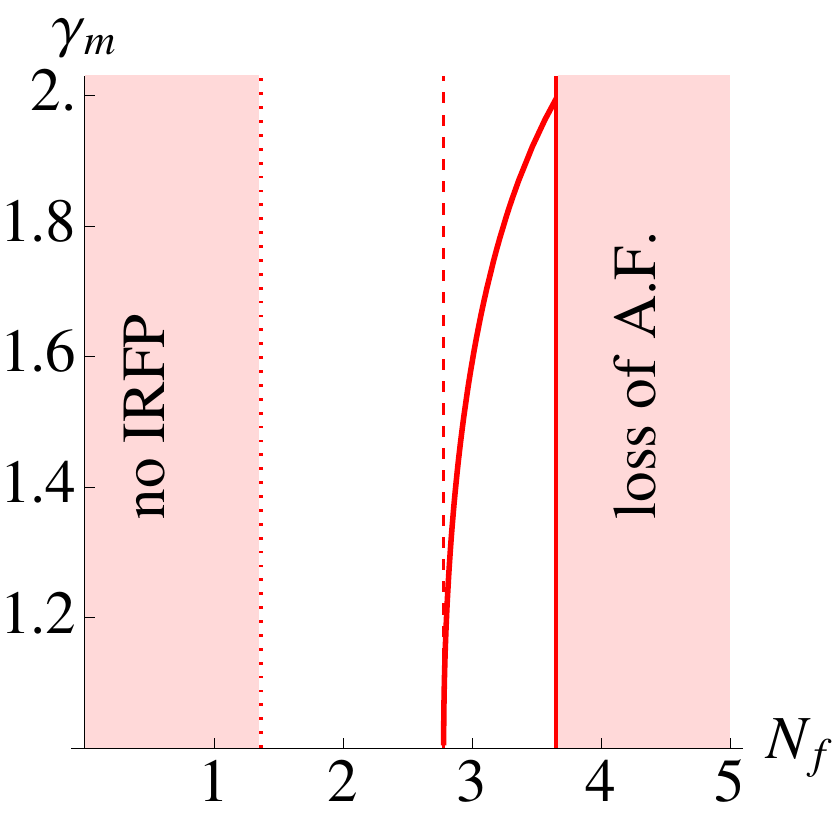} 
\caption{Effect of the four-fermion interactions on the anomalous dimension on the critical line for an $SU(N)$ gauge theories with Dirac fermions transforming according to the two index symmetric representation: From left to right these figures correspond to $N = 3,4$.}
\label{gamma-2symm}
\end{center}
\end{figure}

\begin{figure}[htbp]
\begin{center}
\includegraphics[width=0.3\textwidth,height=0.3\textwidth]{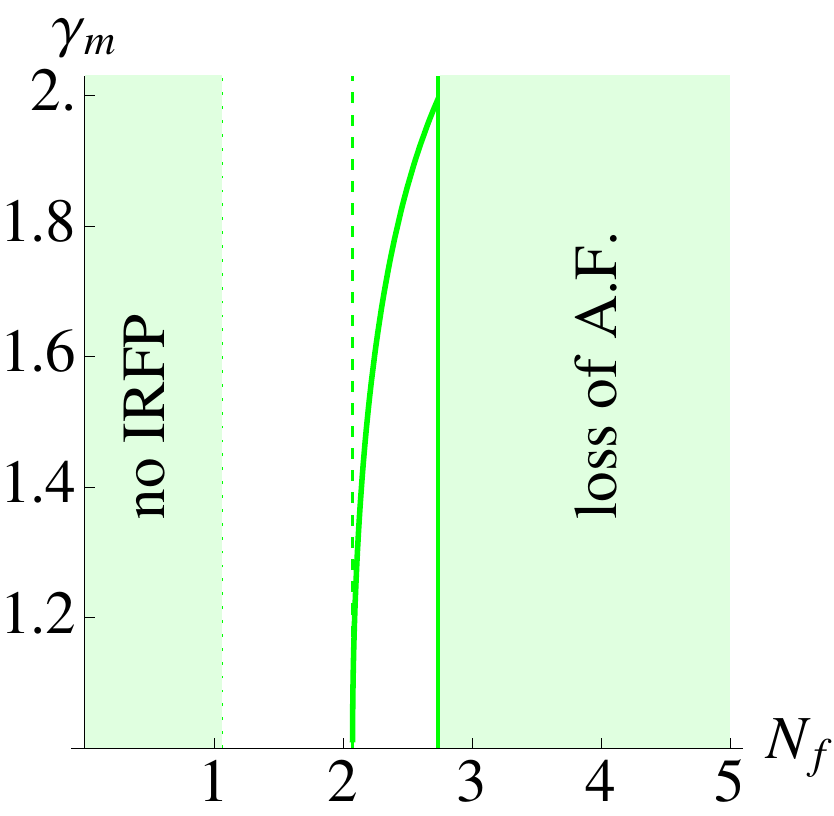} \hspace*{1ex}
\includegraphics[width=0.3\textwidth,height=0.3\textwidth]{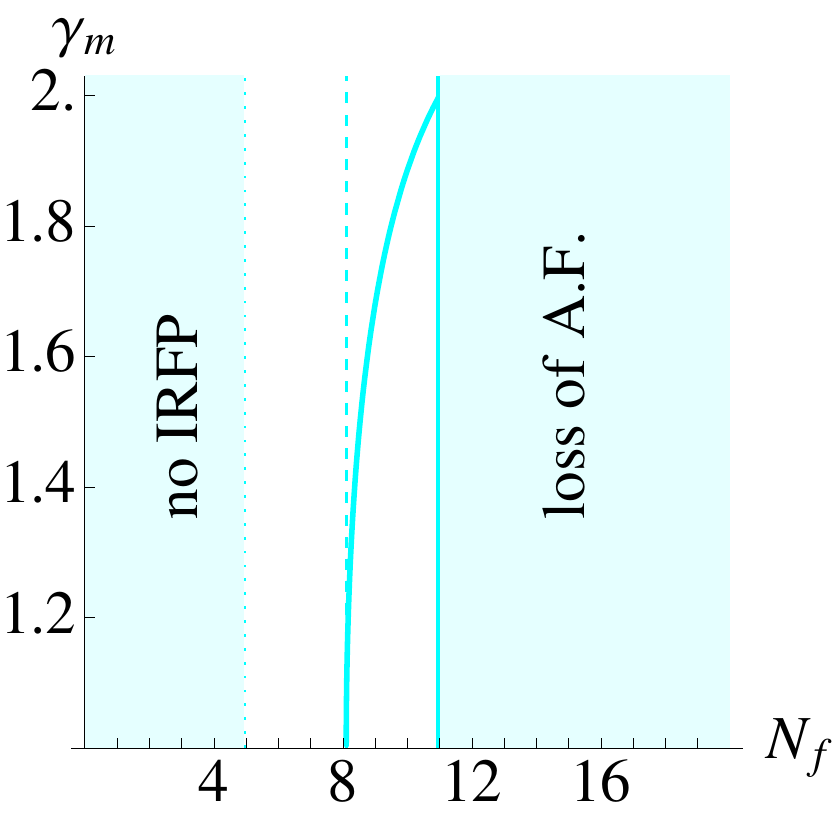}
\caption{Effect of the four-fermion interactions on the anomalous dimension on the critical line for an $SU(N)$ gauge theories with Dirac fermions transforming according to the adjoint and two index anti-symmetric representation: From left to right these figures correspond to the adjoint representation for $N = 2$ and th two index anti-symmetric one for $N = 4$.}
\label{gamma-2asymm-adj}
\end{center}
\end{figure}

The occurrence of large anomalous dimensions of the fermion mass operator $\gamma_m$ in technicolor theories is important to decouple possibly dangerous flavor violation operators from the ones responsible for generating the quark masses \cite{Holdom:1981rm,Yamawaki:1985zg,Appelquist:1986an}, in particular the top. This statement is valid only for certain extensions of the technicolor sector featuring four-fermion interactions at low energy. One can now ask what is the net effect of the four-fermion interactions on the anomalous dimension of the mass. This has been estimated some time ago (see \cite{Yamawaki:1996vr} for a review) and for $0 < \alpha < \alpha^0_{\crit}$ reads \cite{Kondo:1993jq} 
\beq
\gamma_m(g) = 1 -\omega + 2\omega\frac{g}{g_{\crit}}
\quad  \text{where} \quad
\omega = \sqrt{1-\dfrac{\alpha}{\alpha^0_{\crit}}}\,,
\eeq
which is valid both in the broken and unbroken  phase. The anomalous dimension along the critical line ($g = g_{\crit}$ with $1/4 < g < 1$ ) is 
\beq
\gamma_m(g = g_{\crit}) = 1 + \omega \,. 
\eeq
This means that on this line the anomalous dimension is always larger than unity and reaches two for $g=1$.  The dependence on $(N,N_f)$ is estimated using for $\alpha$ in $\omega$ its value at the zero of the two-loop beta function, i.e. $\alpha_*(N,N_f)$, we then have: 
\beq
\gamma_m(N,N_f) = 1 + \omega(N,N_f)\,, 
\eeq
where
\beq
\omega(N,N_f) 
&=& \sqrt{1-\dfrac{\alpha_*(N,N_f)}{\alpha^0_{\crit}}} \nonumber\\[1ex]
&=& \sqrt{ 1 + \dfrac{ 6 C_2(r) \left[ 11 N - 4 N_f C(r) \right]}{17 N^2 -  N_f C(r) \left[10 N+ 6 {C_2(r)}\right]}}\,.
\eeq
Having expressed  $\gamma_m$  as function of the number of flavors, colors and matter representation, at the critical value of $g$, we plot it in Figs.~\ref{gamma-fund},\ref{gamma-2symm} and \ref{gamma-2asymm-adj},  for different phenomenologically relevant models. 
{}For the fundamental representation and $N=2$, $N=3$ and $N=4$ we show $\gamma_m(N,N_f)$ along the critical line in Fig.~\ref{gamma-fund}.  {}For the two index symmetric representation  with $N=3,4$ we plot $\gamma_m(N,N_f)$ in Fig.~\ref{gamma-2symm}. {}For the adjoint of $SU(2)$, left panel, and the two index anti-symmetric of $SU(4)$, right panel, in Fig.~\ref{gamma-2asymm-adj} .
In these figures, the straight dashed line is $N^{\crit}_f$ from equation Eq.(\ref{critf-g}), where the four-fermion interactions are switched off; the solid curve is  $\gamma_m(N,N_f)$ estimated at the critical value of $g$ as function of number of flavors for given representation and number of colors. 
 
{}For each given representation, the anomalous dimension increases monotonously from one to two. The unity value is achieved at the critical number of flavors of the gauge theory without four-fermion interactions. Note also that when the anomalous dimension increases this corresponds also to an increase of the critical coupling $g$ and consequentially at the increase of the critical number of flavors needed to enter the conformal phase. 

\section{The modified diagram}

\begin{figure}[htbp]
\begin{center}
\includegraphics[width=0.49\textwidth]{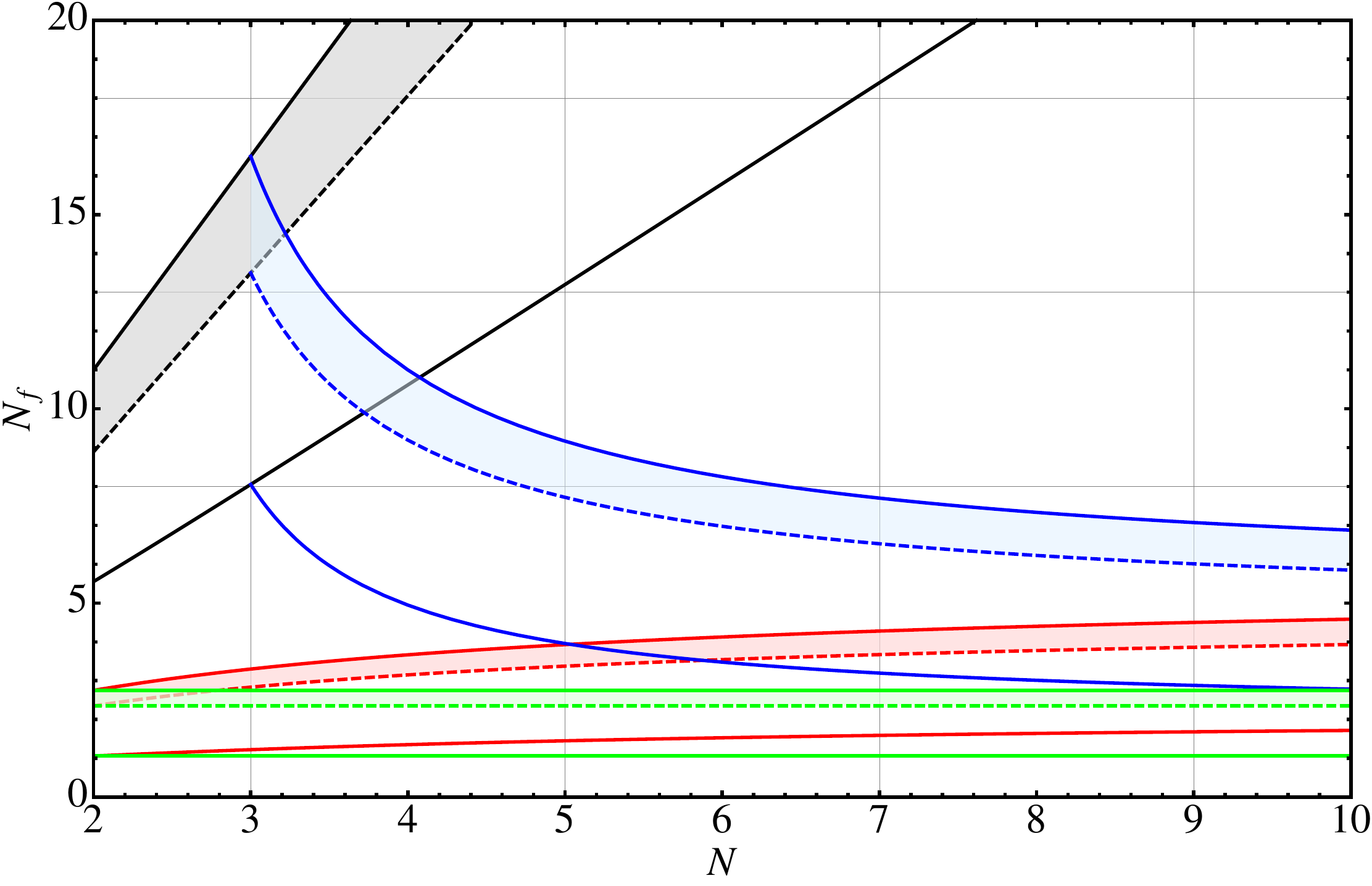} \hspace*{.3ex}
\includegraphics[width=0.49\textwidth]{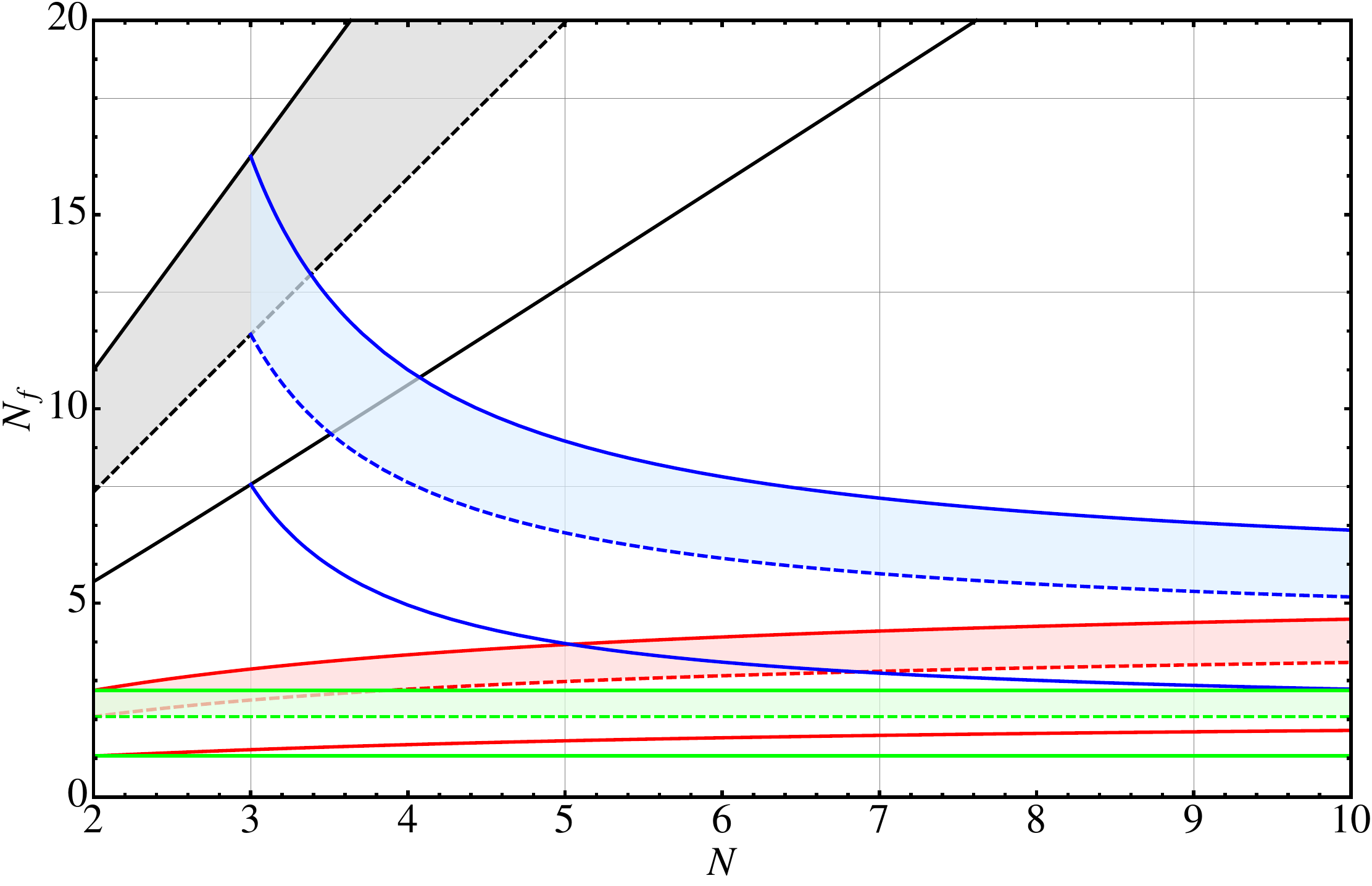}
\caption{Phase diagram for $SU(N)$ gauge theories with (left panel) and without (right panel) four-fermion interactions with Dirac fermions transforming according to different representations of the gauge group.  From top to bottom, we plot the conformal window for theories with fermions transforming according to the: i) fundamental representation (grey), ii) two-index antisymmetric (blue), iii) two-index symmetric (red), iv) adjoint
representation (green) as a function of the number of flavors and
the number of colors. The shaded areas depict the corresponding conformal
windows. The upper solid curve represents loss of
asymptotic freedom,  the lower solid curve indicates when the zero in the two-loops beta function disappear, and  finally above the dashed line, corresponding to $N^{\rm crit}_f[r,g]$ chiral symmetry restores. We have chosen $g=0.75$ for all the dashed lines in the left panel. The right panel is obtained imposing $g=0$ and it is identical to the one obtained in \cite{Sannino:2004qp,Dietrich:2006cm}.}
\label{Phase4}
\end{center}
\end{figure}

It is instructive to plot, see left panel of Fig.\ref{Phase4}, the new phase diagram for a given value of the critical coupling $g$ and matter representation, and then compare it with the one, shown in the right panel, first derived for the two indices representations in \cite{Sannino:2004qp} and then generalized in \cite{Dietrich:2006cm}. {}For consistency both diagrams are obtained using the SD method. However we have shown that the all-order beta function \cite{Ryttov:2007cx}  reproduces, within the uncertainties, the SD result in absence of the four-fermion interactions. We choose, for phenomenological reasons which will become clearer later, the value of $g=0.75$ for which the anomalous dimension is close to two. 
In the left and right panel of Fig.÷\ref{Phase4} we show the phase diagram for the fundamental representation, in black and the upper most window, the two index antisymmetric in blue and the second from the top window, the two index symmetric window is in red and it is the third from the top one, finally in green (straight window) we show the case of the adjoint representation.  
Above the top most solid line, for each representation, the gauge theory looses asymptotic freedom; below the bottom solid line the two-loop beta function does no longer support a zero. In between the top solid line and the dashed line we expect the underlying theory to reach a conformal infrared fixed point. The dashed line in the left panel corresponds to $g = 0.75$ while for the right 
panel $g=0$. The right panel corresponds to the diagram unveiled in \cite{Sannino:2004qp,Dietrich:2006cm}.  

It is clear that the conformal window shrinks implying that, in general, in presence of four-fermion interactions one looses conformality, for a given number of colors and representation, for a higher number of flavors. Interestingly the anomalous dimension increases as well. 

The inclusion of four-fermion interactions will also modify the phase diagram for orthogonal and symplectic gauge theories presented first, using a variety of field analytical methods, in \cite{Sannino:2009aw}. We expect a sizable shrinking of the conformal window in this case too with relevant implications for models of orthogonal technicolor \cite{Frandsen:2009mi}.  Since the phase diagram modifies dramatically it would be very interesting also to analyze the effects of the possible gauge-dual theories emerging when looking at the solutions of the 't Hooft anomaly conditions within the conformal window \cite{Sannino:2009qc,Sannino:2009me}. 

\section{Ideal  Walking Models (IWM)}

Combining the information of the novel phase diagram  together with the knowledge of the anomalous dimension we investigate ideal models of dynamical electroweak symmetry breaking featuring ETC type interactions. Of course, we are using the SD method, and hence the results should be taken {\it cum grano salis}, however our methodology is directly applicable to any other way to determine the conformal window of generic gauge theories with four-fermion interactions. 

We recall some basic features needed for the construction of viable models of dynamical electroweak symmetry breaking featuring traditional type of ETC interactions. 

\begin{itemize}
\item{Non QCD-like dynamics to reduce the contribution to electroweak precision measurements. {}For example,  walking dynamics is expected to reduce the effects of the $S$ parameter \cite{Appelquist:1998xf,Appelquist:1999dq,Duan:2000dy,Kurachi:2006mu,Kurachi:2006ej,Kurachi:2007at}. }

\item{If four-fermion interactions arise due to the presence of the ETC interactions than in order to endow the top quark with the appropriate mass whilst simultaneously reducing potentially dangerous flavor changing neutral currents the anomalous dimension of the techniquarks must be larger than one.}
\end{itemize}
We now scan the new phase diagram, shown in the left panel of Fig.~\ref{Phase4}, in search of ideal extensions of the standard model. It is useful to recall the expressions for the naive $S$-parameter and the top mass:
\begin{eqnarray}
S_{\rm naive} = \frac{N_f}{12\pi} d \left[\rm r\right] \ , \quad m_{\rm Top} =\, \frac{g^2_{\rm ETC}}{\Lambda_{ETC}^2}\, 
\left(\frac{\Lambda_{ETC}}{\Lambda_{TC}}\right)^{\gamma_m }\,\langle \bar{T}T\rangle_{\rm TC} = \frac{g}{\Lambda^2}\,\frac{4\pi^2}{N_f d\left[ r \right]} \left(\frac{\Lambda_{ETC}}{\Lambda_{TC}}\right)^{\gamma_m }\,\langle \bar{T}T\rangle_{\rm TC} \ .
\end{eqnarray}
Here $\Lambda_{TC}\simeq 4\pi F_{TC}$ with $F_{TC}$ of the order of the electroweak condensate, i.e. around $250$~GeV. Of course the precise value of $F_{\pi}$ depends on the specific technicolor model and electroweak embedding. $\Lambda_{ETC}$ is the energy scale at which the ETC interactions generate the four-fermion operator and $\Lambda$ is the gauged NJL scale which we naturally identify with $\Lambda_{ETC}$.  $g_{\rm ETC}$ is the gauge coupling of the underlying ETC interactions. The anomalous dimension is estimated, for the given number of flavors and colors, near the fixed point. We expect the naive $S$-parameter to be an overestimate of the actual value for near walking models \cite{Appelquist:1998xf,Appelquist:1999dq,Duan:2000dy,Kurachi:2006mu,Kurachi:2006ej,Kurachi:2007at} and hence it is a reasonable diagnostic when searching for models of dynamical electroweak symmetry breaking. 
We deduce: 
\begin{equation}
g^2_{\rm ETC} = g \frac{4\pi^2}{N_f d\left[r\right]} \ .
\end{equation}
From the study of the gauged NJL we find that the four-fermion interactions derived from a generic ETC model play a fundamental role on the technicolor dynamics when: 
\begin{eqnarray}
\frac{\pi^2}{N_f d\left[r\right]} <g^2_{\rm ETC} <\frac{4\pi^2}{N_f d\left[ r \right]} \ .
\end{eqnarray}
Given that the extended technicolor interactions are also strongly coupled gauge theories we expect that, in general, it is not possible to decouple their effects on the technicolor dynamics. If, however, the ETC dynamics is less restrictive, i.e. permits for example, a supersymmetric extension or new scalars, then the analysis above  drastically modifies \cite{Antola:2010nt,Antola:2009wq}.  

\subsection{Fundamental Representation: Ideal one family walking model}
The one family walking technicolor model requires the existence of, at least, a new standard model family, including a new neutrino right, to be gauged under the technicolor interactions.  This means that we have a minimum of four doublets of technifermions transforming according to a given representation of the technicolor sector. From the point of view of the technicolor theory one needs eight flavors. In order not to loose asymptotic freedom one is forced to consider these flavors to transform according to the fundamental representation of the underlying gauge group. 

 To reduce the $S$-parameter one chooses the lowest possible number of colors, which for an $SU(N)$ technicolor gauge theory, is $N=2$ yielding $S_{\rm naive}= 4/3\pi$. This value is several standard deviations away from the experimental limits. This model has been considered in \cite{Appelquist:1997fp} and the low energy effective theory featuring the presence of composite massive spin one resonances and topological terms was introduced in  \cite{Appelquist:1999dq,Duan:2000dy}.
 According to several analytic estimates of the conformal window, when the four-fermion interactions are absent, the theory is conformal in the infrared and hence the $S$ parameter vanishes exactly and chiral symmetry does not break. In \cite{Appelquist:1997fp} the authors argue that ordinary strong interactions alone could be sufficient to eliminate the possible infrared fixed point. Below we argue that ETC interactions should be taken into account as well.

In fact, the situation changes dramatically if the ETC interactions are arranged in such a way that the effective coupling is around $g \simeq 0.75$. In this case the critical number of flavors rises to
$8.89$ flavors while the anomalous dimension reaches the value of $\gamma_m \simeq 1.73$ which is large enough to endow the top quark with the physical mass. Remains to be seen if the corrections to the $S$-parameter, which now should also include the effects of the four-fermion interactions, can offset partially or completely the naive large value estimated above. Nevertheless, we have shown that the effects of the four-fermion interactions strongly enhance the phenomenological viability of this interesting model.

\subsection{Higher Representations: Ideal minimal walking }
These models were  conceived to minimize the corrections to the electroweak parameters while still being near conformal \cite{Sannino:2004qp}. See for recent reviews \cite{Sannino:2009za}. 

The simplest {\it Minimal Walking} model consists of an $SU(2)$ gauge theory with two Dirac fermions transforming according to the adjoint representation of the underlying gauge group. We note that this can be also seen as an $SO(3)$ technicolor model, with fermions transforming according to the vector representation. In this case it is straightforward to construct $ETC$ extensions of the model \cite{Evans:2005pu}. The investigation of ETC models for theories with fermions transforming according to higher dimensional representations were also investigated in \cite{Christensen:2005cb}. We have shown \cite{Dietrich:2005jn,Foadi:2007se} that this model satisfies the electroweak constraints because the positive value of the $S$ parameter from the technicolor sector is already much smaller than in the older models and moreover there is a calculable sector constituted by a new lepton doublet (naturally needed to avoid the Witten global anomaly) whose mass spectrum can be arranged to zero this positive contribution. A general effective low energy Lagrangian for this model was presented in \cite{Foadi:2007se} and the first consistent investigations for collider phenomenology appeared in \cite{Belyaev:2008yj}. 

According to SD \cite{Sannino:2004qp} this model is on the verge of developing an infrared fixed point while according to the all-order beta function has already developed an infrared fixed point \cite{Ryttov:2007cx} and the associated anomalous dimension is $\gamma_m = 0.75$. Preliminary lattice results \cite{Catterall:2007yx,Catterall:2008qk,DelDebbio:2008zf,Hietanen:2008vc,Hietanen:2009az,Pica:2009hc,Catterall:2009sb,Lucini:2009an,Bursa:2009we,DelDebbio:2010hu,DelDebbio:2010hx,DeGrand:2009hu,DeGrand:2008kx,DeGrand:2009mt,Fodor:2008hm,Fodor:2009ar,Fodor:2009nh,Kogut:2010cz,Appelquist:2009ty,Appelquist:2009ka,Fodor:2009wk,Fodor:2008hn,Deuzeman:2009mh, Fodor:2009rb,Fodor:2009ff} seem to be consistent with these expectations. We note that, on the lattice it is hard to differentiate a true fixed point from an extreme walking regime. Nevertheless, as it is the case for the fundamental representation, the situation changes dramatically if the ETC interactions are arranged in such a way that the effective coupling is around $g \simeq 0.75$. In this case the critical number of flavors rises to around $2.35$ flavors while the anomalous dimension is $\gamma_m \simeq 1.73$ which is very large and, hence, even better suited to endow the top quark with the physical mass while suppressing flavor changing neutral currents. Differently from the previous case the $S$ was not a problem already at the naive level. A similar analysis can be made for the next to minimal walking technicolor model in which a doublet of technifermions transform according to the two-index symmetric representation of the $SU(3)$ technicolor sector. 
We have, hence, shown that by adding four-fermion interactions, natural in any ETC model, Minimal Walking models became ideal models of dynamical electroweak symmetry breaking.

\section{Conclusions}
We investigated the effects of the presence of four-fermion interactions on the conformal window of gauge theories with fermions transforming according to a given representation of the gauge group. Using the knowledge of the phase diagram of the gauged NJL model we estimated the conformal window again and have shown that it shrinks whilst the anomalous dimension can reach along the lower boundary values which are close to two. 

We discovered that the effects of a strongly coupled ETC sector on technicolor models are welcome and can help healing their problem of the generation of the correct top mass whilst eliminating the tension with flavor changing neutral currents. However, in order to be phenomenologically viable, we should consider models in which the ETC interaction does not commute with the electroweak symmetries \cite{Randall:1992vt,Evans:1994fb} in such a way to be able to give different masses to the top and bottom interactions along the way investigated in \cite{Evans:2005pu} for Minimal Walking models. Models of top-color assisted technicolor are also expected to impact in a similar way on the conformal window analyzed here \cite{Hill:1994hp}. The details of our results will change but, due to the fact, that one will still have four-fermion interactions arising at low energy we expect a similar impact on the conformal window of the associated technicolor theory. 

In the future we would like to explore also the effects on the dynamics and on the conformal window due to the introducing of an explicit mass term for the fermions \cite{Sannino:2008pz,Sannino:2008nv} and the ones of the instantons \cite{Sannino:2008pz}.  The impact on our all-orders beta function \cite{Ryttov:2007cx,Sannino:2009za} and the one modified to take into account the explicit dependence on the mass of the fermions \cite{Dietrich:2009ns} would also be interesting to explore. 

{}First principle lattice simulations can test the new conformal window following the numerical analysis pioneered in \cite{Kocic:1990fq,Hands:1992uv,Kocic:1992vt}

We have shown that the conformal window unveiled here is important for the construction of interesting and more phenomenologically viable models of dynamical electroweak symmetry breaking featuring a strongly coupled ETC sector.

\acknowledgments 
We thank D.D. Dietrich, M.T Frandsen, M.P. Lombardo, E. Pallante, C. Pica, T.~Ryttov and K. Tuominen, for useful discussions.  

\end{document}